\documentclass[12pt, preprint]{emulateapj}
\usepackage{natbib}
\usepackage{mathrsfs}
\usepackage{url}

\newcommand{\kms}{\ifmmode {\rm km~s}^{-1} \else km~s$^{-1}$\fi}
\newcommand{\ergs}{\ifmmode {\rm erg~ s}^{-1} \else erg~s$^{-1}$\fi}
\newcommand{\ergscm}{\ifmmode {\rm erg~s}^{-1} \else erg~s$^{-1}$ cm$^{-2}$\fi}
\newcommand{\Msun}{\ifmmode {\rm M}_{\odot} \else M$_{\odot}$\fi }
\newcommand{\Lsun}{\ifmmode {\rm L}_{\odot} \else L$_{\odot}$\fi}
\newcommand{\qo}{\ifmmode q_{\rm o} \else $q_{\rm o}$\fi}
\newcommand{\Ho}{\ifmmode H_{\rm o} \else $H_{\rm o}$\fi}
\newcommand{\ho}{\ifmmode h_{\rm o} \else $h_{\rm o}$\fi}

\newcommand{\vFWHM}{\ifmmode v_{\mbox{\tiny FWHM}} \else
                    $v_{\mbox{\tiny FWHM}}$\fi}
\newcommand{\CCF}{\ifmmode F_{\it CCF} \else $F_{\it CCF}$\fi}
\newcommand{\ACF}{\ifmmode F_{\it ACF} \else $F_{\it ACF}$\fi}
\newcommand{\Halpha}{\ifmmode {\rm H}\alpha \else H$\alpha$\fi}
\newcommand{\Hbeta}{\ifmmode {\rm H}\beta \else H$\beta$\fi}
\newcommand{\Hgamma}{\ifmmode {\rm H}\gamma \else H$\gamma$\fi}
\newcommand{\Hdelta}{\ifmmode {\rm H}\delta \else H$\delta$\fi}
\newcommand{\Lya}{\ifmmode {\rm Ly}\alpha \else Ly$\alpha$\fi}
\newcommand{\Lyb}{\ifmmode {\rm Ly}\beta \else Ly$\beta$\fi}
\newcommand{\HeI}{\ifmmode {\rm He}\,{\sc i}\,\lambda5876 \else 
	          He\,{\sc i}\,$\lambda5876$\fi}
\newcommand{\HeII}{\ifmmode {\rm He}\,{\sc ii}\,\lambda4686 \else 
	           He\,{\sc ii}\,$\lambda4686$\fi}

\newcommand{\ciii}{\ifmmode {\rm C}\,{\sc iii} \else C\,{\sc iii}\fi}

\newcommand{\oiii}{O\,{\sc iii}}

\def\fake2{\hphantom{3}}

\shorttitle{Narrow-Line Variability}
\shortauthors{Peterson et al.}

\begin{document}

\title{The Size of the Narrow-Line
Emitting Region in the Seyfert 1
Galaxy NGC 5548 from Emission-Line Variability}
\author{B.~M.~Peterson\altaffilmark{1,2},
K.~D.~Denney\altaffilmark{1,3,4,5},
G.~De~Rosa\altaffilmark{1,2},
C.~J.~Grier\altaffilmark{1}, \\
R.~W.~Pogge\altaffilmark{1,2},
M.~C.~Bentz\altaffilmark{6},
C.~S.~Kochanek\altaffilmark{1,2}, 
M.~Vestergaard\altaffilmark{3,7}, \\
E.~Kilerci-Eser\altaffilmark{3},
E.~Dalla~Bont\`{a}\altaffilmark{8,9}, and
S.~Ciroi\altaffilmark{8}
}
\altaffiltext{1}{Department of Astronomy, The Ohio State University,
140 W 18th Ave, Columbus, OH 43210} 
\altaffiltext{2}{Center for
Cosmology \& AstroParticle Physics, The Ohio State University, 191
West Woodruff Ave, Columbus, OH 43210} 
\altaffiltext{3}{Dark Cosmology Centre, Niels Bohr Institute, 
University of Copenhagen,
Juliane Maries Vej 30, DK-2100 Copenhagen, Denmark}
\altaffiltext{4}{Marie Curie Fellow}
\altaffiltext{5}{NSF Astronomy and Astrophysics Postdoctoral Fellow}
\altaffiltext{6}{Department of Physics and Astronomy, Georgia State
University, 25 Park Place, Suite 610,
Atlanta, GA 30303} 
\altaffiltext{7}{Steward Observatory, University of Arizona,
933 North Cherry Avenue, Tucson, AZ  85721}
\altaffiltext{8}{Dipartimento di Fisica e Astronomia `G.\ Galilei',
Universit\`{a} di Padova, Vicolo dell'Osservatorio 3 I--35122, Padova, Italy}
\altaffiltext{9}{INAF--Osservatorio Astronomico di Padova,
Vicolo dell'Osservatorio 5 I--35122, Padova, Italy}
\begin{abstract}
The narrow [O\,{\sc iii}]$\,\lambda\lambda4959,$ 5007 emission-line
fluxes in the spectrum of the well-studied 
Seyfert 1 galaxy NGC 5548 are shown to vary with time.
From this we show that the narrow line-emitting region has
a radius of only 1--3\,pc and is denser
($n_e \sim 10^5\,{\rm cm}^{-3}$) than previously supposed.
The [O\,{\sc iii}] line width is consistent with virial motions
at this radius given previous determinations of the black hole mass.
Since the [O\,{\sc iii}] emission-line flux is usually 
assumed to be constant and is therefore used to calibrate spectroscopic
monitoring data, the variability has ramifications for the long-term
secular variations of continuum and emission-line fluxes,
though it has no effect on shorter-term reverberation studies.
We present corrected optical continuum and broad \Hbeta\ emission-line
light curves for the period 1988 to 2008.

\end{abstract}

\section{INTRODUCTION}
\label{sec:intro}

Beyond the black-hole/accretion disk structure that forms the heart of 
an active galactic nucleus (AGN) and produces the thermal continuum
spectrum, the standard paradigm to explain the 
ultraviolet through infrared spectra of AGNs features two distinct
emission-line regions: a broad-line region (BLR),
which consists of relatively dense
clouds, filaments, or streams of gas deep in the gravitational potential 
well of the central black hole, and a 
lower-density, more spatially extended narrow-line region (NLR)
that gives rise to the forbidden lines (e.g.,
[O\,{\sc iii}]\,$\lambda\lambda 4959$, 5007) and the narrow cores of
the permitted lines. The size of the BLR, $R_{\rm BLR}$,
is measured by reverberation
mapping \citep{Blandford82, Peterson93} and 
is found to be dependent on the 
luminosity of the AGN \citep[][and references therein]{Bentz13} 
roughly as expected from photoionization theory. Recent results
using microlensing in gravitationally lensed quasars, while less
developed, provide an independent confirmation of this picture
\citep[e.g.,][]{Guerras13}. In local Seyfert galaxies,
the size of the BLR is typically light days to light weeks;
even in the nearest AGNs, the BLR projects to only tens of
microarcseconds, so it is currently resolvable only by reverberation mapping
or microlensing. The NLR, by contrast, is sometimes
large enough to be partially resolved
on the sky and often shows a clear biconical structure
\citep[e.g.,][]{Pogge88, Tadhunter89, Schmitt94,
Wilson94, Fischer13}.

The reverberation mapping technique relies on the intrinsic variability
of the AGN continuum and the response of the broad emission lines to these
variations, delayed on average by the
light-travel time across the BLR,
$\tau_{\rm BLR} = R_{\rm BLR}/c$. The continuum and emission-line flux variations
are not extremely large ($\sim10$--20\% on 
broad-line reverberation timescales)
so the reverberation technique requires quite precise
spectrophotometric flux measurements, typically at the 1--2\% level
\citep{Horne04}, to be successful. Unfortunately, 
even under excellent observing conditions, it is difficult to perform
ground-based optical absolute spectrophotometry to better than
$\sim 5\%$ accuracy, so non-standard techniques must be employed to
obtain higher internal precision. It is only necessary to have
accurate {\em relative} flux calibration between epochs, and there are
two standard strategies for doing so.  
The first strategy is to align the spectrograph slit so as to
simultaneously record the spectra of the targeted AGN and a nearby
non-variable star \citep[][and references therein]{Kaspi00}. The second,
more commonly used
method is to assume that the flux in the forbidden 
[O\,{\sc iii}]\,$\lambda\lambda4959$, 5007 lines is constant on 
reverberation time scales and therefore can be used as an internal flux 
calibration standard \citep{Foltz81, Peterson82}.

The assumption that narrow-line fluxes are constant 
over reverberation timescales is justified
by the spatial extent of the NLR.
For case B recombination, the size of the NLR
is of order
\begin{equation}
R_{\rm NLR} \approx 19 \left(\frac{L_{41}({\rm H}\beta)}
{\epsilon n^{2}_{3}} \right)^{1/3}\ {\rm pc},
\end{equation}
where $L_{41}({\rm H}\beta)$ is the luminosity of the H$\beta$ narrow component
in units of $10^{41}\,$ergs\,s$^{-1}$, $n_{3}$ is the electron density
in units of $10^3$\,cm$^{-3}$, 
and $\epsilon$ is the volume filling factor of the narrow-line gas
\citep{Peterson97}. In addition to the large light-travel time across the
NLR, the low particle density means that
the recombination time is also long,
\begin{equation}
\tau_{\rm rec} \approx \left( n_{\rm e} \alpha_{\rm B} \right)^{-1}
\approx 200\,n_3^{-1}\ {\rm years}
\end{equation}
where $\alpha_{\rm B}$ is the hydrogen case B recombination coefficient
and has a value of
$1.43 \times 10^{-13}\,{\rm cm}^{3}\,{\rm s}^{-1}$ at
$T \approx 20,000$\,K \citep{Osterbrock06}.
This effectively damps out any short timescale flux variations.
The luminosity of the \Hbeta\ narrow component in NGC 5548
$(z=0.01718$, heliocentric)
is $\sim 4.3 \times 10^{40}$\,ergs\ s$^{-1}$ (including a correction for
Galactic absorption corresponding to $A_B = 0.088$\,mag), and
NLR electron densities are
typically $n_{\rm e}\approx 2000\,{\rm cm}^{-3}$ \citep{Koski78},
so for NGC 5548, we expect 
$\tau_{\rm NLR} = R_{\rm NLR}/c > 30$\,years and
$\tau_{\rm rec} \approx 100$\,years.

There are, however, only a small number of AGNs that have been spectroscopically
monitored on timescales much longer than BLR reverberation timescales.
In this regard, the Seyfert 1 galaxy NGC\,5548 is a special case. It 
has been the subject of a number of spectroscopic monitoring programs,
motivated at least in part by the desire to have a long time-series on
at least one fairly typical intermediate luminosity AGN
\citep[e.g.,][and references therein]{Sergeev07}; the optical
spectroscopic coverage now extends over 40 years. A long baseline that
includes multiple reverberation campaigns allows us, for example, 
to test the repeatability of black hole mass measurements as the BLR radius
and emission-line profile can change on dynamical timescales
($\tau_{\rm dyn} \approx c\tau_{\rm BLR}/\Delta V$,
where $\Delta V$ is the broad line width)
due to large-scale secular luminosity variations
\citep[e.g.,][]{Peterson02, Gilbert03, KilerciEser13}
that are potentially related to accretion-rate changes
\citep[e.g.,][]{MacLeod10}.

Absolute calibration of the thousands of optical spectra
of NGC 5548 that have been obtained over recent decades
has been tied to a fairly homogeneous subset of spectra that were
obtained under photometric conditions during the first year of the
13-year International AGN Watch monitoring campaign on this source
in 1988--89 \citep{Peterson91}. Based on these observations,
a flux of 
$F([\mbox{O\,{\sc iii}]}\,\lambda5007) = 
5.58 \times 10^{-13}$ ergs\,s$^{-1}$\,cm$^{-2}$
was adopted, and all subsequent spectra were scaled to this value.
For the first seven years of this campaign, 
[O\,{\sc iii}]\,$\lambda5007$ fluxes were measured in a similar high-quality
subset of the spectra \citep{Peterson92, Peterson94, Peterson99,
Korista95} and were found to be consistent with this value
to within the measurement errors, which are estimated to be 
$\sim 4.4\%$ (see \S{\ref{sec:data}}), 
assuming that there is no real variability on time 
scales shorter than a few weeks.

In recent months, we have revisited the issue of absolute flux calibration
of the NGC\,5548 spectra motivated by identification of some apparent
inconsistencies between recent and previously published measurements 
\citep{KilerciEser13} and
by the detection of strong long timescale 
narrow-line flux variability in another AGN in our monitoring program
\citep{DeRosa13}. Here we describe the results
of this re-investigation, which reveals narrow-line flux
variations on timescales short enough to allow determination of the
size of the NLR. We discuss the data and measurements in
\S{2}. We discuss our recalibration of the historical light curve
for NGC 5548 in \S{3} and the implications for the NLR in \S{4}.

\section{DATA AND ANALYSIS}
\label{sec:data}

We use measurements of the absolute
flux of the [O\,{\sc iii}]\,$\lambda5007$ emission line
from high-quality CCD spectra that were obtained through
wide entrance apertures to mitigate seeing
effects \citep[see][]{Peterson95} and under weather 
conditions that were recorded by the observer on site to
be either ``clear'' or ``photometric.'' In each case, the
observing conditions are a judgment call, though the observers appeared to
be quite conservative in declaring weather conditions to be
sufficiently good that absolute spectrophotometry was possible.
For the first 7 years of the International AGN Watch campaign
(1988--89 to 1994--95), we used published
[O\,{\sc iii}]\,$\lambda5007$ emission-line fluxes
\citep{Peterson92, Peterson94, Peterson99, Korista95}.
New measurements were made from International AGN Watch data
from Years 8 through 13 \citep{Peterson02}.
These measurements are given in Table 1.

In addition to the International AGN Watch data, we employed data from
more recent monitoring campaigns that were undertaken by our
group in cooperation with others. We refer to these data sets 
by the internal campaign names we used at MDM Observatory:
AGN05 over 2005 March -- April \citep{Bentz07},
AGN07 over 2007 March -- August \citep{Denney10}, and
AGN12 over 2012 January -- April \citep{DeRosa13}.
We also used data from a campaign undertaken by the
Lick AGN Monitoring Program from 2008 February -- May 
\citep[LAMP08;][]{Bentz09b}. We again selected
a subset of spectra for which the observer noted that the 
weather conditions were good.
Measurements from all of these spectra are given in Table 2.

To estimate the uncertainty in each of the flux measurements in 
Tables 1 and 2, we assumed that the fractional uncertainty is constant for
each set and that there is no true flux variability on the
short timescales. We then compared flux differences between all
pairs of measurements separated by no more than  
10 to 20 days, depending on how many
pairs were available, and assumed that variations on these timescales 
are due only to measurement errors.  The fractional uncertainties 
we adopted are thus 0.044 for the International AGN Watch data,
0.040 for AGN05, 0.065 for AGN07, 0.055 for LAMP08, and
0.028 for AGN12. 

As noted in \S\ref{sec:intro}, 
there are good reasons to believe that the narrow-line
flux should be non-variable on reverberation timescales. Moreover, we
do not see any trend in the [\oiii] fluxes within the individual
data sets that each span less than a year. We therefore
average the flux measurements for each individual data set, and 
these values are given in Table 3. 
We show these average [O\,{\sc iii}]\,$\lambda5007$ fluxes as 
function of Julian Date in Figure 1 (bottom), along with
the 5100\,\AA\ continuum flux (top). There is
a clearly significant long-term downward trend in the 
[O\,{\sc iii}]\,$\lambda5007$ flux as a function of time, that seems to
bottom out and perhaps reverse in the more recent campaigns.

In order to measure directly
the size of the [\oiii]-emitting region and to measure accurately
the width of the [\oiii] lines in a high-resolution spectrum, 
we also retrieved images taken with the {\em Hubble Space Telescope} 
({\em HST}) Space Telescope Imaging Spectrograph (STIS) from the Mikulsky
Archive for Space Telescopes; \cite{Fischer13} have
previously characterized the [\oiii]-emitting region of NGC 5548
as ``compact,'' based on these data.

\section{RECALIBRATION OF NGC 5548 LIGHT CURVES}

Figure 1 shows that our previous assumption 
that a single fixed value of the
[O\,{\sc iii}]\,$\lambda5007$ emission-line flux can serve as the
absolute calibration for all NGC 5548 spectra over long timescales
is incorrect. The [O\,{\sc iii}]\,$\lambda5007$ 
emission-line flux varies significantly on timescales 
as short as a few years.
It is therefore necessary to recalibrate the published
5100\,\AA\ optical continuum and broad H$\beta$ emission-line
fluxes, correcting for the slow variations of the 
[O\,{\sc iii}]\,$\lambda5007$ flux. The flux correction factors
that need to be employed are given in Table 3.
We chose to use average values rather than to fit a 
smooth continuous function of time to all the data because
the differences in the correction factors between closely spaced
campaigns are very small and because we observed no trends within
the individual campaigns.

If we want to place the entire 5100\,\AA\ continuum light curve
on a single flux scale, we must also take into account the different
amounts of host-galaxy starlight contamination in each data set
on account of the different spectrograph entrance apertures employed
in the different observing campaigns. It should be noted that
the International AGN Watch data
sets have all been adjusted relative to a fixed entrance aperture
of $5''\!.0 \times 7''\!.5$.  In Table 4, we give the
host-galaxy fluxes through the various spectrograph entrance
apertures employed in these campaigns based
on modeling the {\em HST} Advanced Camera for Surveys (ACS) 
and Wide Field Camera 3 (WFC3) images
of the host galaxy \citep{Bentz09a, Bentz13}.

For each campaign, we multiply the published 5100\,\AA\
optical continuum light curve by the appropriate correction factor
in Table 3 and then subtract the appropriate
host galaxy flux\footnote{The exception
is the AGN07 data \citep[Table 5 of][]{Denney10}, which
were already corrected for host-galaxy contribution using
an earlier estimate from \cite{Bentz09a}. 
Here we add this earlier estimate back into the fluxes, apply
the flux correction, and then subtract the new host galaxy correction.}
listed in Table 4. The resulting host-corrected light
curve is shown in the top panel of Figure 2 and 
presented in Table \ref{Table:continuum}.

We also recalibrate the long-term emission-line light curve by
multiplying the measurements from each data set by the flux
correction factor from Table 3. We have previously
determined from low-state spectra obtained in Years 2, 4, 9, 12 and 13 that
the flux ratio of narrow \Hbeta\ to [O\,{\sc iii}]\,$\lambda5007$
is constant and has a value of 0.11 \citep{Peterson04}. This ratio
is also the same in spectra from AGN05 and AGN07, so
we conclude that narrow \Hbeta\ is also slowly varying. We therefore
subtract off the narrow \Hbeta\ flux from the total \Hbeta\ flux,
leaving behind the isolated broad component of \Hbeta; this is the 
light curve shown in the lower panel of Figure 2 and presented in 
Table \ref{Table:Hbeta}. We note in passing that the flux ratio 
[\oiii]\,$\lambda5007$/[\oiii]\,$\lambda4959$ is fixed at 2.94
as both lines arise out of the same upper state ($^1D_2$).

The recalibrated fluxes, particularly those obtained in low states,
expose some of the limitations of the simple spectral analysis
employed here. The simple prescription used for the H$\beta$ flux measurement 
\citep[Figure 1 of][]{Peterson91} was intended to capture most of the
\Hbeta\ variations rather than accurate isolate the \Hbeta\ broad-line flux.
For example, according to Table 6, the broad \Hbeta\ flux reaches zero
during the AGN07 campaign (at JD2454251), within the uncertainty
of $\sim4.7\times 10^{-15}$\,ergs\,s$^{-1}$\,cm$^{-2}$. However,
inspection of the original spectrum reveals a clear, but very
weak broad component blended with other weak features. Measurement
of very weak broad-line fluxes can be done accurately only by
employing detailed decomposition and modeling
of the individual spectra, which is work in progress.

\section{IMPLICATIONS FOR THE SIZE AND GAS DENSITY 
OF THE NARROW-LINE REGION}

As with the broad lines, the narrow-line light curve should be a shifted 
and smoothed version of the continuum driving the ionization.  The 
temporal span of our light curves is too short to measure directly the lag
between the continuum and the narrow-line fluxes.  
However, we can statistically
estimate the timescale over which the continuum light curve must be 
smoothed in order to reproduce the narrow-line light curve. This
timescale is roughly equal to the sum of the temporal smoothing
created by the light travel time across the narrow line region 
and the recombination time scale from Equation~2,
\begin{equation}
  \tau_{\rm smooth} \approx  2 R_{\rm NLR} /c + \tau_{\rm rec},
  \label{eqn:tsmooth}
\end{equation}
where we characterize the NLR as a sphere of radius 
$R_{\rm NLR}$.

We first need a statistical model for the continuum variability.  We
obtain this by fitting a damped random walk (DRW) stochastic process
model to the H$\beta$ light curve.  The DRW model is known to well 
reproduce the variability of quasars 
\citep{Kelly09, Kozlowski10, MacLeod10, MacLeod12, Zu13}.
The optical continuum light curve in the most recent campaigns approaches
zero flux in large part because of how the fluxes were measured, so we only
fit the pre-2003 AGN Watch data.  In any case, the large multi-year
gaps in the later data would result in only weak constraints on the
final model fit. We fit the light curve mean and the 
DRW time scale $\tau_{\rm damping}$ 
and amplitude $\sigma$ using the procedures of 
\cite{Kozlowski10} with Monte Carlo Markov Chains (MCMC) to 
estimate the uncertainties.  
Fits to the pseudo-magnitude $-2.5\log F({\rm H}\beta)$
(i.e., fractional variations in flux) had far higher maximum likelihoods
than fits to the flux, 
so we only report
those results, finding   $\log \tau_{\rm damping} = 2.50$, $\log\sigma = -0.31$ 
and $\langle -2.5 \log F({\rm H}\beta)\rangle = -2.00 $ with
90\% confidence ranges of $2.26 < \log\tau_{\rm damping} < 3.04$, 
$-0.33 < \log\sigma < -0.28$ 
and $-2.01 < \langle -2.5\log F({\rm H}\beta) \rangle < -1.95$.  
Here $\tau_{\rm damping}$
is in days and the fluxes in $\sigma$ and $F({\rm H}\beta)$ 
were normalized by $10^{-15}$~ergs$^{-1}$~cm$^{-2}$~\AA$^{-1}$.

We drew model light curves from these MCMC chains, boxcar smoothed them
on timescale $\tau_{\rm smooth}$ and fit them to the seasonally
averaged narrow-line
light curve with a $\chi^2$ statistic.  For the fit, we allowed
a multiplicative scaling between the model of the smoothed 
continuum and the narrow-line data, needed for unit conversion, 
but no temporal shifts.  We assigned the narrow-line data the
epoch of the midpoint of the monitoring campaign.  For each
choice of $\tau_{\rm smooth}$, we made $10^5$ trials and then  
constructed a Bayesian estimate of $P(\tau_{\rm smooth})$ with a logarithmic
prior by weighting each trial as $\exp(-\chi^2/2)$, summing over all
the trials and normalizing the final distribution to unity.
The resulting differential $P(\tau_{\rm smooth})$ and integral
$P(>\tau_{\rm smooth})$ are shown in Figure~\ref{fig:tsmooth}.
The median value is $\tau_{\rm smooth}=14.8$~years  and the 90\% 
confidence range is $7.7$ to $31.3$~years.  
The narrow-line light curve is already smoothed over each
season, with an average length of $0.72$~years.  We can
roughly correct for this by adding the length of the 
average season to $\tau_{\rm smooth}$,  so we adopt as our
final estimate $\tau_{\rm smooth}=15.6$~years with a 90\%
confidence interval of $8.4$ to $32.1$~years. 

Figure~\ref{fig:constrain} shows the constraint this estimate
of $\tau_{\rm smooth}$ places on the size $R_{\rm NLR}$ of the emission region
and the typical density $n_e$.  For high densities, $\tau_{\rm smooth}$
simply becomes the light-crossing time $2R_{\rm NLR}/c$ of the NLR,
and the density is required to be $n_e \gtrsim 10^4$~cm$^{-3}$
to keep $\tau_{\rm rec} < \tau_{\rm smooth}$.  We also show the 
critical density for the [\oiii]\,$\lambda5007$ transition 
\citep[$n^{\rm crit}_{e}\approx 7 \times 10^5$~cm$^{-3}$;][]{Osterbrock06}
and  lines of constant H$\beta$ luminosity 
$L({\rm H}\beta)=4.3 \times 10^{40}$~ergs\,s$^{-1}$
with a filling factor of $\epsilon=1$ and $0.01$, following 
Equation~1. 
\cite{Peterson04} estimate that the mass of the black hole
in NGC 5548 is $M_{\rm BH}  = 7 \times 10^7 M_\odot$. 
\cite{Ferrarese01}
measure the bulge velocity dispersion to be
$\sigma_* = 183 \pm 10$\,km\,s$^{-1}$ which 
together yield a black hole radius of influence 
$R_{\rm BH} = (G M_{\rm BH}/\sigma_*^2) \approx$ 9 pc. Since this is larger than
the estimated NLR radius, the [\oiii] line width of 
$460$\,km\,s$^{-1}$ (see below)
should be dominated by the gravity of the black hole,
with $\Delta V^2 \approx G M_{\rm BH}/R_{\rm NLR}$. 
This leads to a kinematic estimate of the
size of the NLR $R_{\rm NLR} \approx 1.4$\,pc which is
in excellent agreement with all our other estimates,
as shown in Figure \ref{fig:constrain}. It is also noteworthy
that this measurement is in excellent agreement with the
size of the high-ionization component of the NLR predicted
by photoionization equilibrium modeling \citep{Kraemer98}.

The compactness of the [\oiii]-emitting NLR is surprising
given earlier narrow-band imaging that indicated kpc-scale extended
structure \citep{Wilson89}. 
Our own Fabry--Perot [\oiii]\,$\lambda5007$ images
of NGC 5548 \citep{Peterson95} favor a compact NLR, but
these images were obtained under poor seeing ($\sim 2''$)
conditions. We therefore inspected archival STIS images, as described
in \S{\ref{sec:data}}. The [\oiii] 
lines extracted from these data are shown in 
Figure \ref{fig:STIS}. The width of the lines is 
${\rm FWHM} = 7.81\,{\rm \AA} = 460\,{\rm km\,s}^{-1}$. In the spatial
direction, we find that 94\% of the [\oiii] emission is 
concentrated in a 2.42 pixel Gaussian core. 
A TinyTim \citep{Krist11}
model of the point-spread function 
for STIS imaging at the observed wavelength of [\oiii]\,$\lambda5007$
has a width ${\rm FWHM} = 1.38$\, pixels. Subtracting this in quadrature
from the best-fit Gaussian to the [\oiii] emission gives the intrinsic
width of the [\oiii]-emitting region of 2 pixels or 0.102 arcsec or
37\,pc. This means that $\sim94$\% of the [\oiii] emission
arises within a region of 
$R_{\rm NLR} \lesssim 18$\,pc, consistent with the results above.

\section{DISCUSSION AND CONCLUSIONS}

Flux variability and {\em HST} imaging both suggest that the 
[\oiii]-emitting region is much smaller than previously supposed,
1--3\,pc rather than kpc-scale. The actual physical extent of the 
NLR may be somewhat larger than the estimates given
here if it is elongated along the line of sight and if, on account of
obscuration, we preferentially detect emission from the narrow 
line-emitting gas on the
near side. The [\oiii]\,$\lambda\lambda4959,$ 5007 profiles shown in
Figure \ref{fig:STIS} support such a scenario: the lines are distinctly
asymmetric, with ${\rm HWHM}_{\rm blue}/{\rm HWHM}_{\rm red} \approx 2.12$. Such
blueward asymmetries are often assumed to indicate that the NLR has
an outflowing component (e.g., 
\citep[e.g.,][]{Glaspey76a, Glaspey76b, Pelat80, Heckman81, 
Peterson81, Veron81, Whittle85}, which modern spatially resolved
studies have confirmed \citep[e.g.,][]{Storchi-Bergmann12}.
But the outflow velocities in any
case are not terribly different from the virial velocity and none of our
conclusions about the compactness of the NLR would be significantly
altered if the NLR has an outflowing component.

In addition to being compact, we also infer that the particle density
in the [\oiii]-emitting region is higher than previously supposed.
This seems to be consistent with other observations:
it is well known that the widths of the narrow lines in 
Seyfert 2 galaxies correlate with both critical density 
and ionization potential 
\citep{Pelat81, Filippenko84, DeRobertis84, Espey94}. 
Presumably, the higher velocity gas is
closer to the central source, so we can explain this as a consequence of
ionization stratification. From the [\oiii] flux variations,
we infer that the electron density is $\sim 10^5$\,cm$^{-3}$,
close to the critical density.

We suspect that this
is also true for the NLR in other AGNs and there have simply not been
enough high-quality spectrophotometric observations over a sufficiently
long timescale for narrow-line variability to have been reliably 
detected  in many cases. Indeed, we have found only one other 
credible report in the
refereed literature of narrow emission-line flux variability,
the case of 3C\,390.3 \citep{Zheng95}.
This suggests that a systematic search of long-term changes in 
NLR fluxes would be rewarding and that, in principle, the
time variability of the NLR would provide an independent check
of black hole mass estimates from reverberation mapping of the BLR.

The narrow-line flux variations that we report on here do not
negate any of the conclusions to date that have been drawn from
reverberation mapping studies. The narrow-line fluxes are still
effectively constant on timescales much longer than reverberation timescales.
Indeed, if the narrow-line gas is virialized, as the broad-line
gas seems to be, then 
$\tau_{\rm NLR}/ \tau_{\rm BLR} \approx  
(\Delta V[{\rm broad}]/\Delta V [{\rm narrow}])^2$,
which is of order
100. These results do, however, show that absolute calibration
of the narrow-line fluxes requires attention.

\acknowledgments We are grateful to the National Science Foundation
for support of this work through grants
AST-1008882 (BMP, GDR, CJG, and RWP) and
AST-1009756 (CSK) 
to The Ohio State University,
and PHY11-25915 (MV) to the Kavli Institute for Theoretical Physics, 
a CAREER Grant AST-1253702 (MCB) to Georgia State University,
and a Postdoctoral Research Fellowship 
AST-1302093 (KDD).
CJG is grateful for support through an
Presidential Fellowship from The Ohio State University.
EDB is supported by Padua University through grants
60A02-1283/10, 60A02-5052/11, and 60A02-4807/12.
MV thanks the Kavli Institute for Theoretical Physics at University
of California, Santa Barbara, for their hospitality where part of this work was
done. The research leading to these results has received funding from the
People Programme (Marie Curie Actions) of the European Union's Seventh
Framework Programme FP7/2007-2013/ under REA grant agreement No. 300553 (KDD).
The Dark Cosmology Centre is funded by the Danish National Research
Foundation.  This research has made use of
the NASA/IPAC Extragalactic Database (NED), which is operated by the
Jet Propulsion Laboratory, California Institute of Technology, under
contract with the National Aeronautics and Space Administration.

\clearpage

\raggedright

\begin{deluxetable}{llc}
\tablewidth{0pt}
\tablecaption{Unpublished AGN Watch Measurements}
\tablehead{
\colhead{UT Date} &
\colhead{File Name} &
\colhead{$F(\mbox{[O\,{\sc iii}]}\,\lambda5007$)} \\
\colhead{(1)} &
\colhead{(2)} &
\colhead{(3)} 
} 
\startdata
1995 Nov 27 & n50052h &  $  5.21 \pm    0.23  $ \\
1996 Jan 11 & n50093a &  $  5.03 \pm    0.22  $ \\
1996 Apr 02 & n50175a &  $  5.30 \pm    0.23  $ \\
1996 Jun 28 & n50262a &  $  4.93 \pm    0.22  $ \\
1996 Sep 10 & n50336h &  $  4.76 \pm    0.21  $ \\
1997 Feb 14 & n50493a &  $  5.14 \pm    0.23  $ \\
1997 Jun 24 & n50623a &  $  4.88 \pm    0.22  $ \\
1997 Jul 06 & n50635a &  $  5.10 \pm    0.22  $ \\
1997 Aug 04 & n50664h &  $  5.20 \pm    0.23  $ \\
1997 Sep 06 & n50697h &  $  5.49 \pm    0.24  $ \\
1998 Jan 25 & n50838a &  $  5.16 \pm    0.23  $ \\
1998 Mar 03 & n50875a &  $  4.94 \pm    0.22  $ \\
1998 May 22 & n50955a &  $  5.02 \pm    0.22  $ \\
1998 Jul 23 & n51017h &  $  5.00 \pm    0.22  $ \\
1998 Aug 31 & n51056h &  $  5.52 \pm    0.24  $ \\
1998 Sep 17 & n51077h &  $  4.44 \pm    0.20  $ \\
1999 Jan 10 & n51189h &  $  4.67 \pm    0.20  $ \\
1999 Feb 12 & n51221h &  $  4.66 \pm    0.20  $ \\
1999 Feb 23 & n51233h &  $  5.27 \pm    0.23  $ \\
1999 Mar 12 & n51250h &  $  5.08 \pm    0.22  $ \\
1999 Apr 24 & n51293h &  $  5.06 \pm    0.22  $ \\
1999 Jul 09 & n51368h &  $  5.10 \pm    0.22  $ \\
1999 Jul 18 & n51377h &  $  5.11 \pm    0.22  $ \\
1999 Aug 17 & n51407h &  $  5.44 \pm    0.23  $ \\
1999 Sep 10 & n51431h &  $  4.81 \pm    0.21  $ \\
2000 Dec 21 & n51900h &  $  4.82 \pm    0.21  $\\
\enddata
\label{Table:AGNWatch}
\tablecomments{Column (1) gives the UT Date of the observation.
Column (2) gives the name of the spectrum as it appears in
the International AGN Watch 
archive (see http://www.astronomy.ohio-state.edu/$\sim$agnwatch),
where the prefix ``n5'' refers to the galaxy, the following
digits are the four least-significant figures in the Julian Date of
observation, and the final letter indicates the origin of the
data \citep[see Table 2 of][]{Peterson02}. The measured observed frame
[O\,{\sc iii}]\,$\lambda5007$ flux appears in column (3), in
units of $10^{-13}$\,ergs\,s$^{-1}$\,cm$^{-2}$.}
\end{deluxetable} 

\begin{deluxetable}{lcc}
\tablewidth{0pt}
\tablecaption{Measurements from Recent Monitoring Campaigns}
\tablehead{
\colhead{JD} &
\colhead{} &
\colhead{ } \\
\colhead{(2400000+)} &
\colhead{$F(\mbox{[O\,{\sc iii}]}\,\lambda5007$)} &
\colhead{Reference} \\
\colhead{(1)} &
\colhead{(2)} &
\colhead{(3)} 
} 
\startdata
53438.0 &  $    4.52 \pm     0.18$&  1 \\
53438.9 &  $    4.68 \pm     0.19$&  1 \\
53447.0 &  $    4.75 \pm     0.19$&  1 \\
53460.9 &  $    4.42 \pm     0.18$&  1 \\
53464.6 &  $    4.49 \pm     0.18$&  1 \\
53465.9 &  $    4.51 \pm     0.18$&  1 \\
53466.9 &  $    4.36 \pm     0.18$&  1 \\
53469.9 &  $    4.41 \pm     0.18$&  1 \\
53470.9 &  $    4.11 \pm     0.16$&  1 \\
53471.9 &  $    4.58 \pm     0.18$&  1 \\
54191.8 &  $    4.92 \pm     0.32$&  2 \\
54201.8 &  $    4.73 \pm     0.31$&  2 \\
54204.8 &  $    4.54 \pm     0.29$&  2 \\
54205.8 &  $    4.71 \pm     0.31$&  2 \\
54212.8 &  $    4.46 \pm     0.30$&  2 \\
54215.8 &  $    4.98 \pm     0.32$&  2 \\
54223.8 &  $    4.96 \pm     0.32$&  2 \\
54230.8 &  $    4.43 \pm     0.29$&  2 \\
54231.8 &  $    5.10 \pm     0.33$&  2 \\
54236.8 &  $    4.63 \pm     0.30$&  2 \\
54239.8 &  $    4.56 \pm     0.30$&  2 \\
54245.8 &  $    4.57 \pm     0.30$&  2 \\
54248.8 &  $    4.68 \pm     0.30$&  2 \\
54250.8 &  $    4.50 \pm     0.29$&  2 \\
54255.8 &  $    4.25 \pm     0.28$&  2 \\
54258.8 &  $    4.38 \pm     0.28$&  2 \\
54260.8 &  $    5.11 \pm     0.33$&  2 \\
54261.8 &  $    4.70 \pm     0.30$&  2 \\
54264.8 &  $    4.49 \pm     0.29$&  2 \\
54265.8 &  $    3.88 \pm     0.25$&  2 \\
54566.9 &  $    5.27 \pm     0.29$&  3 \\
54566.9 &  $    5.18 \pm     0.29$&  3 \\
54568.8 &  $    4.17 \pm     0.23$&  3 \\
54568.8 &  $    4.14 \pm     0.23$&  3 \\
54569.9 &  $    4.16 \pm     0.23$&  3 \\
54569.9 &  $    4.25 \pm     0.23$&  3 \\
54587.9 &  $    4.29 \pm     0.24$&  3 \\
54587.9 &  $    4.24 \pm     0.23$&  3 \\
54588.9 &  $    4.23 \pm     0.23$&  3 \\
54588.9 &  $    4.28 \pm     0.24$&  3 \\
54589.9 &  $    4.38 \pm     0.24$&  3 \\
54589.9 &  $    4.38 \pm     0.24$&  3 \\
54590.9 &  $    4.29 \pm     0.24$&  3 \\
54590.9 &  $    4.26 \pm     0.23$&  3 \\
54596.9 &  $    4.43 \pm     0.24$&  3 \\
54596.9 &  $    4.43 \pm     0.24$&  3 \\
54597.9 &  $    4.30 \pm     0.24$&  3 \\
54597.9 &  $    4.29 \pm     0.24$&  3 \\
54604.9 &  $    3.99 \pm     0.22$&  3 \\
54604.9 &  $    3.62 \pm     0.20$&  3 \\
55932.9 &  $    4.70 \pm     0.13$&  4 \\
55935.9 &  $    4.63 \pm     0.13$&  4 \\
55936.9 &  $    4.74 \pm     0.13$&  4 \\
55940.6 &  $    5.00 \pm     0.14$&  4 \\
55940.9 &  $    4.77 \pm     0.13$&  4 \\
55942.6 &  $    5.03 \pm     0.14$&  4 \\
55945.9 &  $    4.76 \pm     0.13$&  4 \\
55946.9 &  $    4.91 \pm     0.14$&  4 \\
55960.9 &  $    4.59 \pm     0.13$&  4 \\
55979.6 &  $    4.71 \pm     0.13$&  4 \\
55980.6 &  $    4.53 \pm     0.13$&  4 \\
55983.6 &  $    4.95 \pm     0.14$&  4 \\
55987.5 &  $    4.91 \pm     0.14$&  4 \\
56001.9 &  $    4.68 \pm     0.13$&  4 \\
56002.9 &  $    4.72 \pm     0.13$&  4 \\
56007.9 &  $    4.78 \pm     0.13$&  4 \\
56009.9 &  $    4.82 \pm     0.13$&  4 \\
56010.9 &  $    4.71 \pm     0.13$&  4 \\
56014.9 &  $    4.70 \pm     0.13$&  4 \\
56015.9 &  $    4.98 \pm     0.14$&  4 \\
56017.9 &  $    4.67 \pm     0.13$&  4 \\
56025.9 &  $    4.72 \pm     0.13$&  4 \\
\enddata  
\tablecomments{Column (1) gives the Julian Date of observation.
Column (2) gives the observed frame
[O\,{\sc iii}]\,$\lambda5007$ emission line
flux and its estimated uncertainty
in units of $10^{-13}$\,ergs\,s$^{-1}$\,cm$^{-2}$.
Column (3) gives the reference for
each spectrum. References:
(1) AGN05 \citep{Bentz07}; 
(2) AGN07 \citep{Denney10};
(3) LAMP08 \citep{Bentz09b};
(4) AGN12 \citep{DeRosa13}.}

\end{deluxetable} 

\begin{deluxetable}{lcccc}
\tablewidth{0pt}
\tablecaption{Adopted Mean [O\,{\sc iii}]\,$\lambda5007$ 
Fluxes and Correction Factors}
\tablehead{
\colhead{ } &
\colhead{ } &
\colhead{ } &
\colhead{ } &
\colhead{Flux} \\
\colhead{ } &
\colhead{ } &
\colhead{JD Range} &
\colhead{} &
\colhead{Correction}\\
\colhead{Data Set} &
\colhead{Ref.} &
\colhead{(2400000+)} &
\colhead{$\langle F(\mbox{[O\,{\sc iii}]}\,\lambda5007) \rangle$} &
\colhead{Factor} \\
\colhead{(1)} &
\colhead{(2)} &
\colhead{(3)} &
\colhead{(4)} &
\colhead{(5)} 
} 
\startdata
AGN Watch, Yr 1 (1988--89)   & 1 & 47509 -- 47809 & $5.586 \pm 0.278$ & 1.001 \\
AGN Watch, Yr 2 (1989--90)   & 2 & 47861 -- 48179 & $5.484 \pm 0.239$ & 0.983 \\
AGN Watch, Yr 3 (1990--91)   & 3 & 48225 -- 48534 & $5.396 \pm 0.165$ & 0.967 \\
AGN Watch, Yr 4 (1991--92)   & 3 & 48623 -- 48898 & $5.519 \pm 0.280$ & 0.989 \\
AGN Watch, Yr 5 (1992--93)   & 4 & 48954 -- 49256 & $5.620 \pm 0.172$ & 1.007 \\
AGN Watch, Yr 6 (1993--94)   & 5 & 49309 -- 49637 & $5.355 \pm 0.517$ & 0.960 \\
AGN Watch, Yr 7 (1994--95)   & 5 & 49679 -- 50008 & $5.386 \pm 0.125$ & 0.965 \\
AGN Watch, Yr 8 (1995--96)   & 6 & 50044 -- 50373 & $5.163 \pm 0.346$ & 0.925 \\
AGN Watch, Yr 9 (1996--97)   & 6 & 50435 -- 50729 & $5.162 \pm 0.220$ & 0.925 \\
AGN Watch, Yr 10 (1997--98)  & 6 & 50775 -- 51085 & $5.013 \pm 0.350$ & 0.898 \\
AGN Watch, Yr 11 (1998--99)  & 6 & 51142 -- 51456 & $5.022 \pm 0.263$ & 0.900 \\
AGN Watch, Yr 12 (1999--2000)& 6 & 51517 -- 51791 & $\ldots$          & 0.882 \\
AGN Watch, Yr 13 (2000--01)  & 6 & 51879 -- 52265 & $4.820 \pm 0.212$ & 0.864 \\
AGN05 \citep{Bentz07}        & 6 & 53431 -- 53472 & $4.485 \pm 0.178$ & 0.804 \\
AGN07 \citep{Denney10}       & 6 & 54180 -- 54333 & $4.629 \pm 0.298$ & 0.830 \\
LAMP08 \citep{Bentz09b}      & 6 & 54509 -- 54617 & $4.330 \pm 0.353$ & 0.776 \\
AGN12 \citep{DeRosa13}       & 6 & 55932 -- 56048 & $4.772 \pm 0.136$ & 0.855\\
\enddata
\tablecomments{The monitoring campaign name and reference are given
in columns (1) and (2) and the range of Julian Dates is given in 
columnn (3).  The average and standard deviation
[O\,{\sc iii}]\,$\lambda5007$ emission-line fluxes
are given in column (4) in units of $10^{-13}$\,ergs\,s$^{-1}$\,cm$^{-2}$.
Column (5) is the
ratio of the flux given in column (4) to the 
previously adopted absolute flux of 
$F(\mbox{[O\,{\sc iii}]}\,\lambda5007) = 
5.58 \times 10^{-13}$ ergs\,s$^{-1}$\,cm$^{-2}$.
References:
(1) \cite{Peterson91};
(2) \cite{Peterson92};
(3) \cite{Peterson94};
(4) \cite{Korista95};
(5) \cite{Peterson99};
(6) This work.}
\label{Table:cfactors}
\end{deluxetable} 

\begin{deluxetable}{lcc}
\tablewidth{0pt}
\tablecaption{Host Galaxy Contributions}
\tablehead{
\colhead{Data Set} &
\colhead{Aperture Geometry (arcsec)} &
\colhead{$F_{\rm gal}[{\rm 5100\,\AA}\, (1+z)]$} \\
\colhead{(1)} &
\colhead{(2)} &
\colhead{(3)} 
} 
\startdata
AGN Watch, Years 1--13 & $5 \times  7.5 $ & $3.75 \pm 0.38$ \\
AGN05                  & $5 \times 12.75$ & $4.34 \pm 0.43$ \\
AGN07                  & $5 \times 12.0 $ & $4.27 \pm 0.43$ \\
LAMP08                 & $4 \times 9.4$   & $3.54 \pm 0.35$ \\
AGN12                  & $5 \times 15.0$  & $4.45 \pm 0.44$ \\ 
\enddata
\tablecomments{Column (1) identifies the individual monitoring campaigns.
Column (2) gives the nominal spectrograph entrance aperture
(projected slit width and extraction window)
used in each campaign.
Column (3) gives the adopted 5100\,\AA\ observed frame
host-galaxy flux through that aperture,
based on the nucleus-free model of the NGC 5548 host galaxy
from \cite{Bentz13}, in 
units of $10^{-15}$\,ergs\,s$^{-1}$\,cm$^{-2}\,{\rm \AA}^{-1}$.}
\label{Table:GalaxyFlux}
\end{deluxetable} 

\clearpage

\begin{deluxetable}{lc}
\tablewidth{0pt}
\tablecaption{Revised Continuum Light Curve}
\tablehead{
\colhead{Julian Date} &
\colhead{$F_\lambda(5100\,{\rm \AA}[1 +z])$}\\
\colhead{(1)} &
\colhead{(2)}
} 
\startdata
 47509.000 & $  5.239 \pm    0.360 $ \\
 47512.000 & $  5.990 \pm    0.731 $ \\
 47517.000 & $  5.970 \pm    0.390 $ \\
 47524.000 & $  6.050 \pm    0.390 $ \\
 47525.000 & $  6.370 \pm    0.400 $ \\
 47528.000 & $  6.821 \pm    0.420 $ \\
 47533.000 & $  6.400 \pm    0.410 $ \\
 47534.000 & $  6.811 \pm    0.420 $ \\
 47535.000 & $  6.420 \pm    0.410 $ \\
 47539.000 & $  7.001 \pm    0.380 $ \\
\enddata
\tablecomments{A complete version of Table 5 is
available in machine-readable form.
Column (1) is the Julian Date $-2400000$.
Column (2) gives the observed-frame AGN flux
at rest wavelength 5100\,\AA, corrected for the host galaxy
contribution, in units of 
$10^{-15}$\,ergs\,s$^{-1}$\,cm$^{-2}\,{\rm \AA}^{-1}$.}
\label{Table:continuum}
\end{deluxetable} 
\begin{deluxetable}{lc}
\tablewidth{0pt}
\tablecaption{Revised H$\beta$ Emission-Line Light Curve}
\tablehead{
\colhead{Julian Date} &
\colhead{$F(\Hbeta)$}\\
\colhead{(1)} &
\colhead{(2)}
} 
\startdata
 47509.000 & $     6.924 \pm    0.260 $ \\ 
 47512.000 & $     7.634 \pm    0.450 $ \\ 
 47517.000 & $     7.424 \pm    0.280 $ \\ 
 47524.000 & $     8.135 \pm    0.310 $ \\ 
 47525.000 & $     8.025 \pm    0.300 $ \\ 
 47528.000 & $     7.704 \pm    0.290 $ \\ 
 47530.000 & $     7.484 \pm    0.280 $ \\ 
 47533.000 & $     7.915 \pm    0.300 $ \\ 
 47534.000 & $     7.654 \pm    0.290 $ \\ 
 47535.000 & $     7.764 \pm    0.290 $    
\enddata
\tablecomments{A complete version of Table 6 is
available in machine-readable form.
Column (1) is the Julian Date $-2400000$.
Column (2) gives the observed-frame \Hbeta\ emission-line
flux, corrected for the \Hbeta\ narrow component, in units of
$10^{-13}$\,ergs\,s$^{-1}$\,cm$^{-2}$.}
\label{Table:Hbeta}
\end{deluxetable} 


\clearpage
\begin{figure}
\begin{center}
\plotone{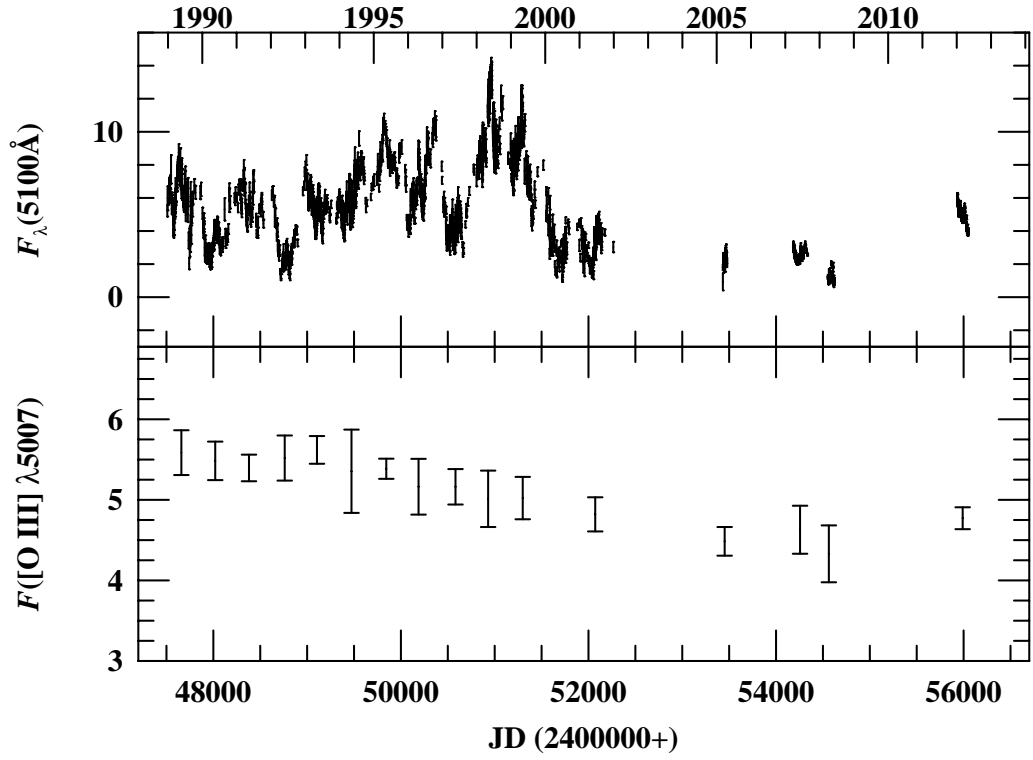}
\caption{Top panel: Observed frame
5100\,\AA\ AGN continuum flux as a function of time, calibrated to a constant
value of the [O\,{\sc iii}]\,$\lambda 5007$ flux with a host-galaxy
contribution (Table 4) removed,
in units of $10^{-15}$ ergs\,$^{-1}$\,cm$^{-2}$\,\AA$^{-1}$.
Original sources of data are cited in
the notes to Table 3. Bottom panel: Observed frame time-averaged 
[O\,{\sc iii}]$\,\lambda5007$ fluxes measured from spectra taken on 
nights that observers recorded night-sky conditions to be
``clear'' or ``photometric,''
in units of $10^{-13}$ ergs\,$^{-1}$\,cm$^{-2}$, as given in
Table 3.}
\label{fig:lightcurves}
\end{center}
\end{figure}

\begin{figure}
\begin{center}
\plotone{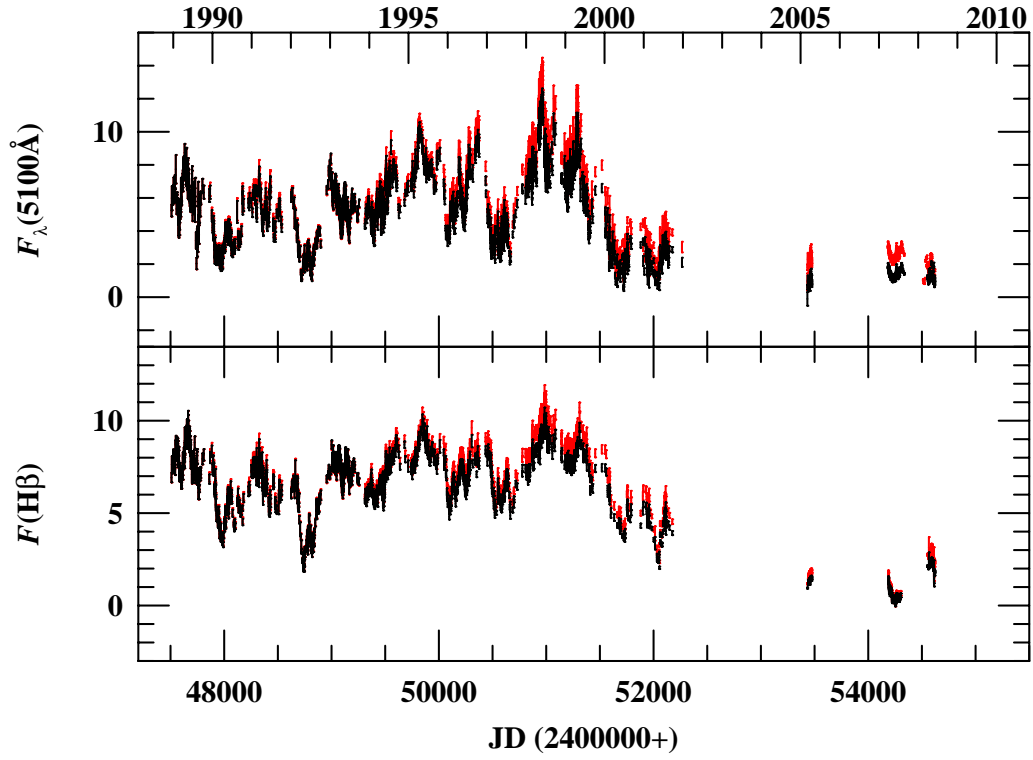}
\caption{Recalibrated observed-frame AGN continuum (top panel,
in units of $10^{-15}$ ergs\,$^{-1}$\,cm$^{-2}$\,\AA$^{-1}$)
and broad H$\beta$ emission-line 
(bottom panel, 
in units of $10^{-13}$ ergs\,$^{-1}$\,cm$^{-2}$)
light curves for NGC\,5548. 
The lower axis shows the Julian Date and the upper axis
shows calendar year. The light curves are flux calibrated to
the [O\,{\sc iii}]$\,\lambda5007$ fluxes given in Table 3 and shown in
Figure 1. The continuum has been corrected for the host galaxy
contribution as given in Table 4, and a narrow H$\beta$ contribution
has been subtracted from the broad H$\beta$ light curve
assuming a narrow line flux 11\% that of 
[O\,{\sc iii}]\,$\lambda 5007$. The original light curves are plotted in
red, and the recalibrated light curves are plotted over them
in black.}
\label{fig:lccompare}
\end{center}
\end{figure}

\begin{figure}
\plotone{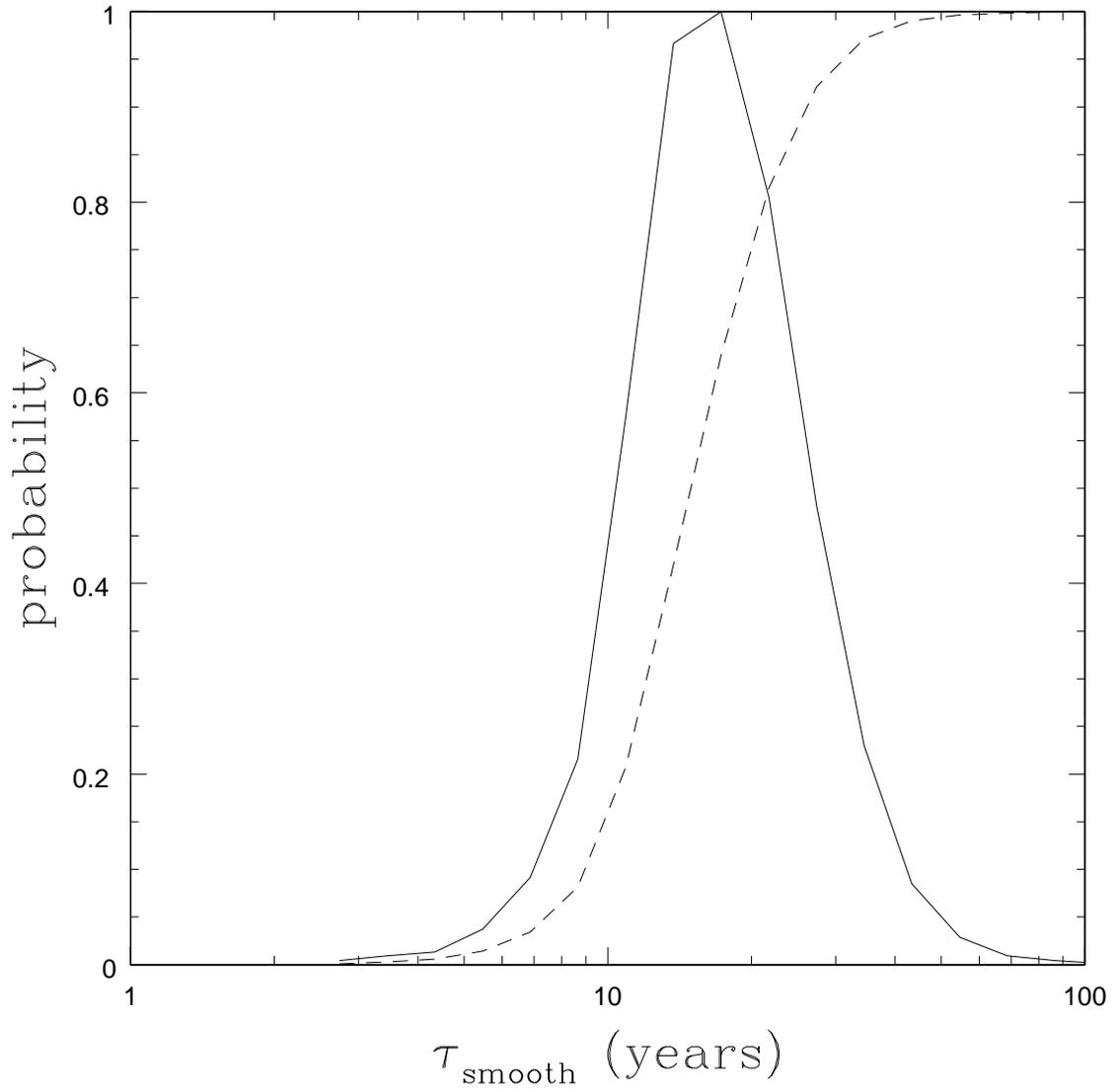}
\caption{
  Differential (solid) and integral (dashed) probability 
  distributions for the smoothing time scale $\tau_{\rm smooth}$ 
  required to match the continuum variability to the narrow
  line variability.  The differential probability distribution
  is normalized to have a peak of unity.  This does not include
  the correction for the seasonal averaging of the
  [\oiii] fluxes.
  }
\label{fig:tsmooth}
\end{figure}

\begin{figure}
\plotone{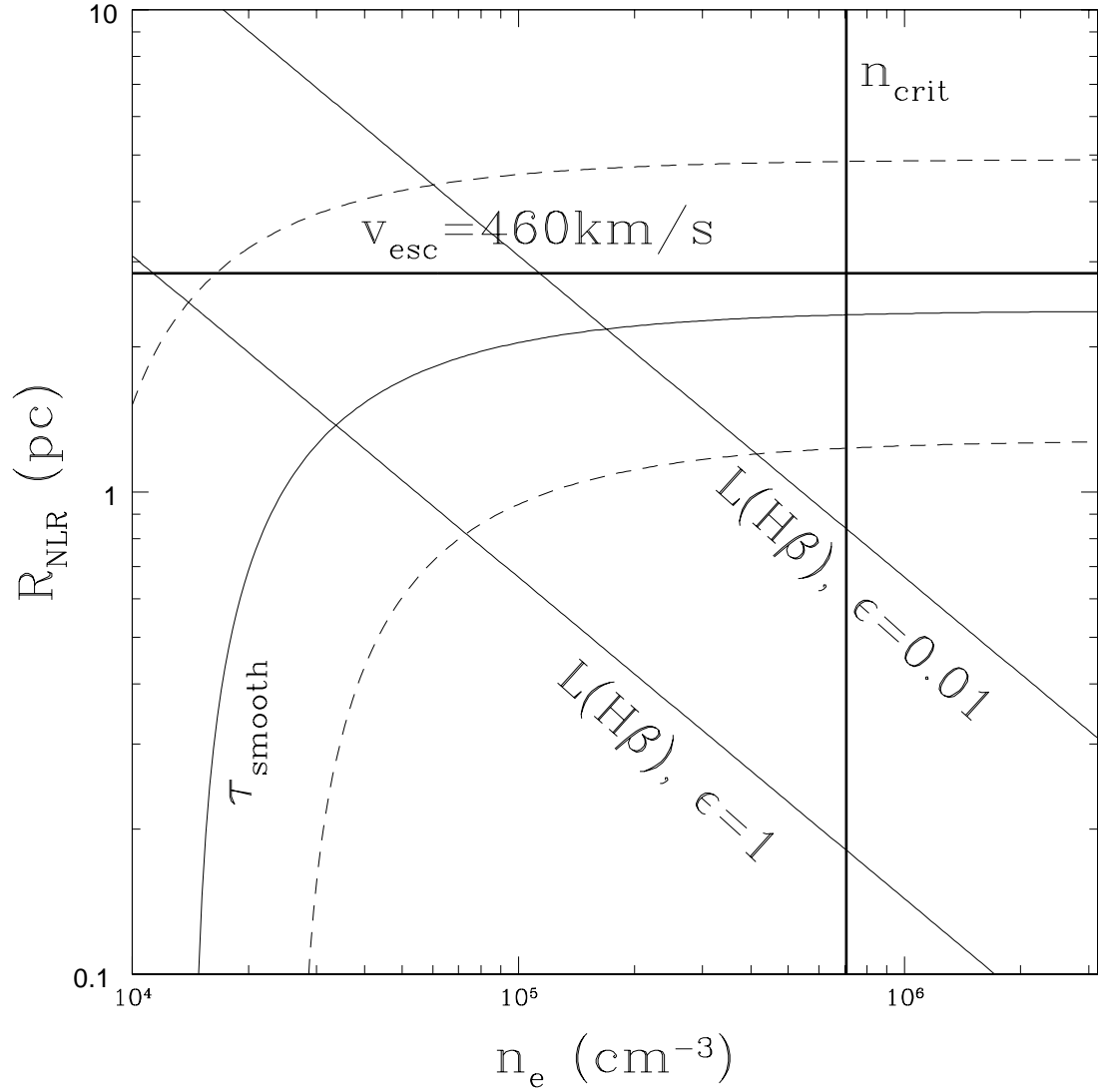}
\caption{
  Constraints on the narrow line region density $n_e$ and
  size $R_{\rm NLR}$.  The curves labeled $\tau_{\rm smooth}$ show the
  median (solid) and 90\% confidence limits (dashed) from
  the observed narrow-line variability.  The lines labeled
  $L({\rm H}\beta)$ show the conditions needed to produce the
  observed H$\beta$ luminosity for a filling factor of
  either $\epsilon=1$ or $0.01$.  The heavy vertical line
  indicates the critical density for [\oiii] emission.
  The heavy horizontal line indicates the radius where
  the escape velocity is $v_{\rm esc}=460$\,\kms\ for 
  $M_{\rm BH} = 7 \times 10^7 M_\odot$.
  }
\label{fig:constrain}
\end{figure}

\begin{figure}
\plotone{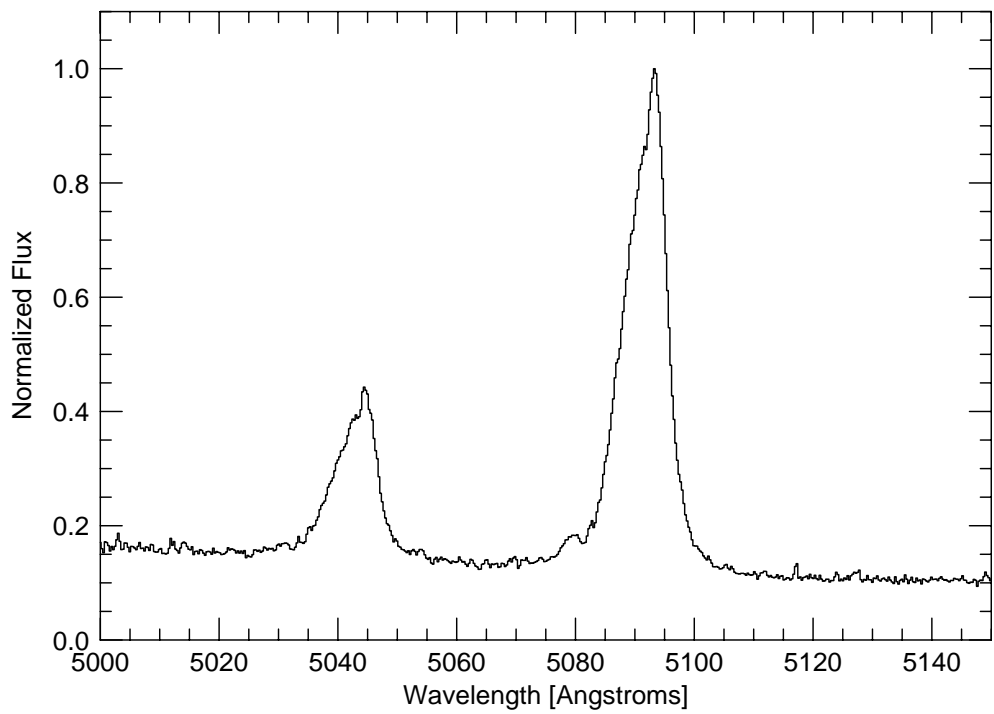}
\caption{
{\em HST} STIS spectrum of the [\oiii]\,$\lambda\lambda 4959$, 5007
lines in NGC 5548.
  }
\label{fig:STIS}
\end{figure}

\end{document}